# TIME EVOLUTION OF SU(1,1) COHERENT STATES

J. Zaleśny

Institute of Physics, Technical University of Szczecin

Al. Piastów 48, 70-310 Szczecin, Poland

Mathematical aspects of the SU(1,1) group parameter $\xi$ dynamics governed by Hamiltonians exhibiting some special types of time dependence has been presented on an elementary level from the point of view of Möbius transformation of complex plane. The trajectories of $\xi$ in continuous and mappings in discrete dynamics are considered. Some simple examples have been examined. Analytical considerations and numerical results have been given.

PACS numbers: 03.65.Fd, 02.20.Sv

# 1. INTRODUCTION

A general method for constructing coherent states for an arbitrary Lie group has been given by Perelomov [12,13]. In this paper we are interesting in time evolution of the SU(1,1) coherent states. Various aspects of the dynamics were examined by many authors. Some relevant examples concerning continuos dynamics the reader may find in articles [1,2,4,7,8,9,11] and discrete dynamics in [3,5,6]. For a review paper on construction and classification of coherent states see [15]. The dynamics driven by a SU(1,1) Hamiltonian has been much investigated mainly in the context of the two-photon processes, the generation of squeezed states of light by nonlinear optical processes (e.g. degenerate parametric amplifiers or down-converters), the propagation of an electromagnetic wave in a nonhomogeneous medium with a quadratic dependence of the refractive index on the transverse coordinates, and time-dependent harmonic oscillators. The reader may find many references in the review article [14]. The evolution of quantum states ruled by the time-dependent coherence-preserving SU(1,1) Hamiltonian can be analyzed using the Wei-Norman method which allows the possibility of representing the time-evolution operator as a finite product of exponential operators, where each exponent contains a product of a group generator and a time-dependent complex function [14]. The functions obey a system of Riccati type nonlinear differential equations. Solutions of them cannot be obtained in general.

In this paper we do not use the Wei-Norman method. Our attention is focused on the SU(1,1) group parameter $\xi$. Nevertheless the equation of motion obeyed by this parameter is also a Riccati type equation. Because its general solutions are unknown, in order to gain the idea about the SU(1,1) dynamics we examine some interesting examples of special time-dependence. Both continuous and discrete dynamics is studied and connections between the two cases are shown. As far as we know, the discrete dynamics was considered in the case of infinitely narrow δ-like function pulses [3,6,10]. For instance Gerry *et al*. [6] use modulated real δ-like pulsing function and Bechler et al.[3] consider so called 'kicked dynamics' with non-modulated complex coupling parameter. In this paper a model with modulated pulses of *finite* width and magnitude is used, which is maybe more realistic assumption in comparison to the 'kicked-like' dynamics models. Though our treatment is 'classical' rather than quantum but in the considered case of the coherence-preserving Hamiltonian (1) it is equivalent to the genuine quantum mechanical problem.

The paper is organized as follows. In section 2 we introduce briefly the idea of SU(1,1) coherent states, and the equation of motion of the SU(1,1) group parameter $\xi$. In section 3 we investigate some of the most characteristic features of the continuous dynamics based upon the equation of motion for $\xi$. We try to classify the types of trajectories of the SU(1,1) group parameter $\xi$ on the phase space obtained for various frequencies of coupling parameter. We give here some numerical and analytical results. In section 4 we construct iteration equation for the discrete values of the parameter $\xi_n$. The evolution is described by Poincaré-type evolution maps. We give some numerical examples of the maps for various pulsing functions. We note here that all 'stroboscopic equations' in case of SU(1,1) group have the form of Möbius automorphism of the unit circle. At the end we show how the chain of Möbius transformations may be replaced by the chain of linear transformations of the complex plane. We use this approach to express in non-time-dependent pulsing case the *n* step of iteration via the initial condition. The last section 5 contains summary of the paper and a very brief discussion of Lapunov exponent and the question of chaos in SU(1,1) systems.

## 2. THE MODEL

We consider a model described by the Hamiltonian given as a *linear* combination of the SU(1,1) group generators with time-dependent coefficients

$$H = 2\hbar\omega K_0 + \hbar\chi(t)K_+ + \hbar\overline{\chi}(t)K_- \quad . \tag{1}$$

The generators obey the following rules of commutation

$$[K_0, K_\pm] = \pm K_\pm \quad , \quad [K_-, K_+] = 2K_0 \tag{2}$$

The bar over symbols means complex conjugation.

A Schwinger-Wigner-type realization of SU(1,1) can be given in terms of (more familiar to physicists) harmonic-oscillator creation and annihilation operators

$$K_0 = \tfrac{1}{4}(a^+ a + a a^+) \quad , \quad K_+ = \tfrac{1}{2}(a^+)^2 \quad , \quad K_- = \tfrac{1}{2}a^2 \tag{3}$$

Because the Hamiltonian is linear in the generators, the coherent character of the generalized coherent states associated with the noncompact Lie group SU(1,1) under time evolution is preserved, which means that the quantum and classical evolutions are essentially identical [13, 4]. The generalized SU(1,1) coherent states $|\xi\rangle$ are defined and constructed in the way first proposed by Perelomov [12, 13]. They are characterized by the complex parameter $\xi$ for which $0 \leq |\xi| < 1$. We shall use the SU(1,1) group parameter $\xi$ as representation of the phase space. The 'classical' equation of motion for $\xi$ is [4]

$$\dot{\xi} = \{\xi, H\} \quad , \tag{4}$$

where { , } is the Poisson bracket defined as

$$\{A, B\} = \frac{(1 - |\xi|^2)^2}{2ik}\left(\frac{\partial A}{\partial \xi}\frac{\partial B}{\partial \overline{\xi}} - \frac{\partial A}{\partial \overline{\xi}}\frac{\partial B}{\partial \xi}\right) , \tag{5}$$

and

$$H = \langle \xi | H(K_0, K_\pm, t) | \xi \rangle \quad . \tag{6}$$

The constant $k$ in (5) is the Bargmann index and one might take it as $k = {}^1\!/_4$ here.

The resulting equation from (4) is

$$\dot{\xi} = -2i\omega\xi - i\overline{\chi}\xi^2 - i\chi \tag{7}$$

As we have equivalence of classical and quantum description of the dynamics in the sense that parameter $\xi$ exactly follows the quantum state $|\xi\rangle$, we can restrict our attention to the motion of the point $\xi$ in the unit circle on the complex plane. It remains true also in the discrete case. Another derivation of equation (7) the reader may find in [3, 11].

In this paper we examine dynamics in continuous and discrete case followed from eq. (7) assuming that $\chi$ is time-dependent.

## 3. CONTINUOUS APPROACH

### 3.1 Formulation of the problem

The phase-space for solutions of the equation (7) is a unit circle in the complex plane. Because of time-dependent coefficient $\chi(t)$ the equation belongs to the class of nonautonomic differential equations. In general $\chi(t)$ is complex and may be written in form $\chi(t) = c(t)\, e^{-if(t)}$, where $c(t)$, $f(t)$ are real functions. However, further, we restrict our considerations to less general form, when $f(t) = k'\omega t$ ($k'$ is a real number). We can come to a rotating frame with frequency $2\omega$ in order to eliminate this frequency from the motion. It is equivalent to use the interaction picture. So we put $k'=2+k$ and our choice for $\chi(t)$ is

$$\chi(t) = c(t)e^{-i(2+k)\omega t} \qquad (8)$$

We seek solution of (7) in the form

$$\xi(t) = a(t)e^{-i(2+k)\omega t}, \qquad (9)$$

where unknown complex function $a(t)$ describes motion in the rotating frame. It is also limited to the interior of the unit circle i.e. $0 \le |a(t)| < 1$. Substituting (9) into (7) we obtain

$$\dot{a}(t) = i\omega k a(t) - ic(t)\left[a^2(t) + 1\right] \qquad (10)$$

This equation is some case of the Riccati equation. A general solution of it is unknown. To investigate it we consider some simple cases.

## 3.2 Some exact solutions

For $k = 0$ ($k' = 2$) the solution can be found explicitly. We have in this case

$$a(t) = i\frac{1 + e^{2S(t)}(a_0 - i)/(a_0 + i)}{1 - e^{2S(t)}(a_0 - i)/(a_0 + i)}, \quad \text{where} \quad S(t) = \int_0^t c(\tau)d\tau, \quad (11)$$

and $a_0$ is the initial value of $a(t)$. In particular, for $a_0 = 0$ we get

$$\xi(t) = -i\, th(S(t))e^{-2i\omega t} \quad (12)$$

The behavior of $\xi(t)$ depends on function $S(t)$. When $S(t)$ fulfills the condition

$$\lim_{t \to \infty} |S(t)| \to \infty, \quad (13)$$

then $\xi(t)$ forms more or less regular spiral, i.e. the trajectory comes up closer and closer to the unit circle, without however reaching it. Further we will use the name 'spiral' to describe such noncompact trajectory, inessential how much the curve resembles an ordinary, regular spiral. Arising of spiral in this case is a very characteristic feature for frequency $2\omega$, independent of the initial conditions, see Fig.1 as an example.

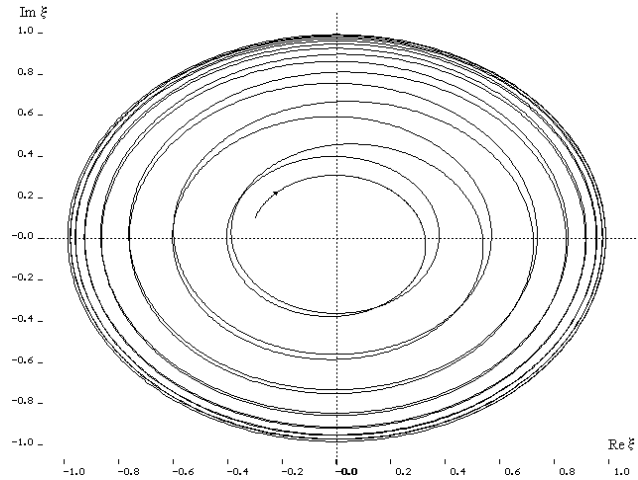

Fig.1: $2\omega$ frequency case. Parameters: $c(t) = 0.1\sin t + 0.05$; $\omega = 1$; $k = 0$; $\xi_0 = -0.3 + 0.1i$.

The spiral is formed because of putting the movement $a(t)$ toward the circumference on the rotation with frequency $2\omega$. One of the most regular spiral can be obtained for constant $c$ and $a_0 = 0$. The point $a(t)$ runs from the point (0,0) along imaginary axis of the complex plane to the circumference of the unit circle. The time dependence of $a(t)$ is given by the function −

th(*ct*), which for small *t* is simply − *ct*. Thus, for small *t* in ξ picture it is so called Archimedes spiral.

Quite different behavior we can observe if *S*(*t*) does not fulfill the condition (13). For instance, if $a_0 = 0$ and values of *S*(*t*) are limited to an interval, in ξ(*t*)-picture we observe more or less (it depends on complexity of function *c*(*t*) ) complicated figures drawn by point ξ(*t*) on complex plane, e.g. see Fig.2.

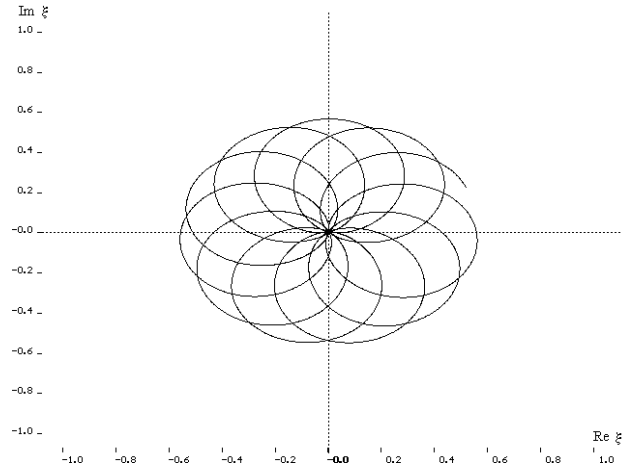

Fig.2: 2ω frequency case. Parameters: $c(t) = \sin \pi t$; $\omega = 1$; $k = 0$; $\xi_0 = 0$.

Note that in this case the trajectory remains in an area of radius less then a unit. This is true also for other initial values. Further we will use the name 'compact figures' to describe behavior like this.

Another exact solution of eq.(10) can be found for arbitrary parameter *k* and constant *c*. It is

$$a(t) = \frac{k_1}{2} \frac{1 - (k_2/k_1) e^{-iS(t)\sqrt{\Delta}} (2a_0 - k_1)/(2a_0 - k_2)}{1 - e^{-iS(t)\sqrt{\Delta}} (2a_0 - k_1)/(2a_0 - k_2)} , \quad (14)$$

where $k_1 = \frac{1}{\alpha} + \sqrt{\Delta}$ , $k_2 = \frac{1}{\alpha} - \sqrt{\Delta}$ , $\Delta = \frac{1}{\alpha^2} - 4$ , $\alpha = \frac{c}{k\omega}$ , $S(t) = ct$ .

For an initial $a_0 = x_0 + iy_0$ the trajectory of the point $a(t) = x(t) + iy(t)$ is given as

$$(x - A)^2 + y^2 = B^2 \quad \text{where:} \quad A = \frac{x_0^2 + y_0^2 - 1}{2x_0 - \frac{1}{\alpha}} , \quad B^2 = A^2 + 1 - \frac{A}{\alpha} . \quad (15)$$

Spiral solutions appear only when circle (15) has crossover points with unit circle. Then, one of them is an attracting fixed point, and the other one is repulsing. The point $a(t)$ runs along the circle (15) but it cannot achieve the attracting point in finite time. On the maps for $\xi(t)$ we observe the spiral, because of the rotating term in eq.(9). The coordinates of the fix points are given by

$$x = \frac{1}{2\alpha} \quad , \quad y = \pm\sqrt{1 - \frac{1}{4\alpha^2}} \tag{16}$$

The obvious condition $|x| < 1$ gives us important inequality

$$|k| < |2c/\omega| \tag{17}$$

For a given constant amplitude $c$ it determines the rotational frequency $k'\omega$ for which spiral solutions appear. E.g., for $k = 0$, i.e. in frequency $2\omega$ case, the inequality is valid for any nonzero $c$. For $k = -2$, i.e. zero frequency, it follows that $|\omega| < |c|$.

For various initial values, eq.(15) gives us a whole map of trajectories in rotating frame, which are nonconcentric circles with centers lying on the real axis. For $|\alpha| < \frac{1}{2}$ all the circles are entirely inside the unit circle. There are also two elliptic fixed points on the real axis, but only one of them, that lying inside the unit circle is interesting for us (Fig.3).

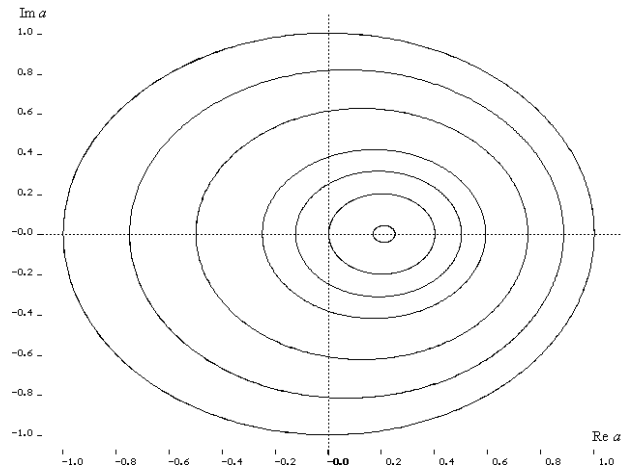

Fig.3: The family of trajectories in the rotating frame. Parameters: $c = 0.2$; $\omega = 1$; $k = 1$; nine initial values on real axis: $a_0 = -1, ... , 1$.

The lager is $|\alpha|$ the closer is the elliptic fixed point to the boundary of the unit circle. For $|\alpha|=1/2$ the fixed point achieves the unit circle and changes its character becoming parabolic fixed point. For lager $|\alpha|$ the parabolic fixed point 'splits' into two hyperbolic fixed points, attractor and repulser (Fig.4).

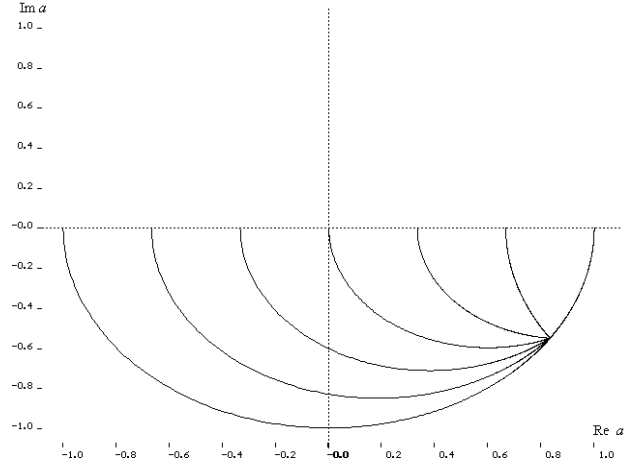

Fig.4: The family of trajectories in the rotating frame. Parameters: $c = 0.6$; $\omega =1$; $k =1$; seven initial values on real axis: $a_0 = -1, ... ,1$.

At the end of this point we examine some linear approximation of eq.(10). This approximation may be used near the center of the unit circle, where $|a|<<1$. Instead of eq. (10) one may use

$$\dot{a}(t) = i\omega k a(t) - ic(t) \ . \qquad (18)$$

A general solution of this equation can be easily found. In particular, for the most natural in this approximation initial value $a_0 =0$ the solution (including also the rotating term) is

$$\xi(t) = -iR(k,t)e^{-2i\omega t} \qquad \text{where} \qquad R(k,t) = \int_0^t c(\tau)e^{-ik\omega\tau}d\tau \ . \qquad (19)$$

Note that R($k=0$, $t$)=S($t$). For $a_0=0$ and constant $c$ the trajectory obtained in the linear approximation is the same as that corresponding to the exact solution (the circle (15) ). The difference between exact and approximated solutions appears only in time dependence of the movement along the trajectory.

## 3.3 Arising of spirals

Undoubtedly we should expect arising a spiral if $|\xi(t)|=|a(t)|\to 1$ for $t\to\infty$. Taking eq.(10) and equation conjugated to it, one can obtain equation of motion for $|a(t)|$ as

$$\frac{1}{2}\frac{d}{dt}\left(|a(t)|^2\right) = -c(t)\operatorname{Im} a(t)\left(1-|a(t)|^2\right). \tag{20}$$

This equation is $k$ independent, so it have to be valid for all $k$. The formal solution of it is

$$1-|a(t)|^2 = \left(1-|a_0|^2\right)e^{2T(t)} \quad , \text{ where } \quad T(t) = \int_0^t c(\tau)\operatorname{Im} a(\tau)d\tau. \tag{21}$$

The spiral-like solution occurs if $T(\infty)\to -\infty$. The most obvious way in which the condition can be fulfilled is

$$c(t)\operatorname{Im} a(t) < 0, \quad \text{for every } t. \tag{22}$$

The simple example is the case of $c$ constant in value but changing its sign every time when the point $a(t)$ (rotating with frequency $\omega_c = c\sqrt{\Delta} = k\omega\sqrt{1-4\alpha^2}$) crosses the $x$-axe. It results in spiral behavior even for $|\alpha| < \tfrac{1}{2}$.

The condition (22) is also fulfilled in the $2\omega$ frequency case. It is interesting however that some effective $2\omega$ frequencies can be introduced. E.g., let the coefficient $\chi(t)$ is given as

$$\chi(t) = c_0 \sin(\kappa\omega t)e^{-ik'\omega t} \quad , \text{ where } c_0 \text{ is constant.} \tag{23}$$

The $\sin(\kappa\omega t)$ twice changes sign in its period $\tau$, i.e. there is $2\pi$ phase gain in time interval $\tau$. During this time the vector $e^{-ik'\omega t}$ turns by an angle $k'\omega\tau$. The total angle $\alpha = k'\omega\tau + 2\pi$, so effective frequency $\omega_{eff} \equiv \alpha/\tau = k'\omega + 2\pi/\tau$, i.e., $\omega_{eff} = (k' + \kappa)\omega$.

Note that we can think about $2\pi$ phase not as a gain but as a loss. It leads to effective frequency $\omega_{eff} = (k' - \kappa)\omega$. If we take $k'$ and $\kappa$ in the following way

$$k' + \kappa = 2 \quad \text{or} \quad k' - \kappa = 2, \quad \text{i.e.} \quad \kappa = -k \quad \text{or} \quad \kappa = k, \tag{24}$$

then we have $2\omega$ effective frequency case. Moreover the condition (13) is fulfilled. Indeed, in numerical experiments we observe spirals similar to that one shown in Fig.1.

### 3.4 Constant $\chi$

For completeness we give the solution of eq. (7) for constant $\chi$. It will be used later to construct the iteration equation in a discrete case. For initial value $\xi(0) = \xi_0$ the solution is

$$\xi(t) = \frac{P(p-\omega) + (p+\omega)e^{-2ipt}}{\overline{\chi}(P - e^{-2ipt})} \qquad (25)$$

where: $\qquad p = \sqrt{\omega^2 - |\chi|^2} \quad , \qquad P = \frac{\overline{\chi}\xi_0 + \omega + p}{\overline{\chi}\xi_0 + \omega - p} \quad .$

For real $p$ the trajectories of eq. (25) are non-concentric circles. There is one elliptic fixed point inside the unit circle. For imaginary $p$ two hyperbolic fixed points, one stable and one unstable appear. Both lying on the unit circle. In another way the result (25) may be obtained from eq.(14), if $k=-2$ and $\chi=c$. The shape of trajectories for real $p$ and imaginary $p$ is exactly the same as those shown in Fig.3 and Fig.4. Certainly, at present they represent trajectories of point $\xi(t)$. For any complex $\chi = ce^{i\beta}$ we need only to rotate the above picture given for $c$ about angle $\beta$.

At last, it is worth to mention that only for $\omega > |\chi|$ the dynamics is well defined, since only then the Hamiltonian (1) is bounded from below [11].

## 4. DISCRETE APPROACH

### 4.1 Formulation of the problem

Our discrete model is as follows. We divide time on segments of length $T$. In each segment evolution from $(n-1)T$ to $nT - t$ is free (i.e. $\chi = 0$) and described by

$$\xi(\tau) = \xi_0 e^{-2i\omega(\tau)} \qquad (26)$$

where $T > t$ and $\xi_0$ is constant in a segment.

For time from $nT - t$ to $nT$ we impose constant $\chi$ different from zero, so evolution is governed by the eq. (25). Different values of $\chi_n$ can be in different segments. We examine discrete values of $\xi_n$ just after the pulses. As a result the pulsing dynamics is described by the iteration equation

$$\xi_n = \frac{A_n \xi_n + B_n}{\overline{B}_n \xi_n + \overline{A}_n} , \qquad (27)$$

where
$$A_n = [p_n \cos(p_n t) - i\omega \sin(p_n t)] e^{-i\omega(T-t)}$$
$$B_n = -i\chi_n \sin(p_n t) e^{i\omega(T-t)} \qquad p_n = \sqrt{\omega^2 - |\chi_n|^2}$$

This form of $A_n$, $B_n$ is valid for real $p_n$ ($\omega > \chi$). If one uses imaginary $p_n$ e.g. in 'kicked dynamics' then trigonometric functions change into hyperbolic functions.

Formula (27) is a special case of so called Möbius transformations well known in the complex plane theory. These transformations form a group. Since our phase space is limited to the interior of the unit circle it is enough to examine a subgroup of all Möbius transformations, that map unit circle into itself, i.e., automorphisms of the unit circle

$$z_{n+1} = e^{i\theta^{(n)}} \frac{z_n - a^{(n)}}{1 - \overline{a}^{(n)} z_n} . \qquad (28)$$

The transformation depends on one real parameter $\theta^{(n)}$ and one complex parameter $a^{(n)}$ for which $|a^{(n)}| < 1$. The upper index $n$ indicates that the parameters may be step-dependent. Eq.(27) is a special case of eq.(28) and the group parameters are determined by

$$e^{i\theta^{(n)}} = \frac{A_n}{\overline{A}_n} , \quad a^{(n)} = -\frac{B_n}{A_n} . \qquad (29)$$

The form (28) is very general. It embraces the models of pulsed SU(1,1) dynamics discussed previously in literature, e.g. [3, 6]. Even continuous dynamics may be treated as a special case of it, if one puts $T = t$ and $T \to 0$.

The group property of the automorphism enables in principle to write down the form of the solution after $n$ steps via the initial condition $z_0$.

$$z_n = e^{i\theta_n}\frac{z_0 - a_n}{1 - \bar{a}_n z_0} \quad , \tag{30}$$

where $a_n$, $\theta_n$ may be called as effective parameters. Note that they have lower indices in contrary to the current parameters $\theta^{(n)}$, $a^{(n)}$. It is easy to find iteration equations for effective parameters

$$e^{i\theta_{n+1}} = e^{i\theta^{(n+1)}}\frac{e^{i\theta_n} + a^{(n+1)}\bar{a}_n}{1 + a_n\bar{a}^{(n+1)}e^{i\theta_n}} \quad , \qquad a_{n+1} = \frac{a^{(n+1)} + a_n e^{i\theta_n}}{e^{i\theta_n} + a^{(n+1)}\bar{a}_n} \quad . \tag{31}$$

From eqs. (30) and (31) we note that if $|z_n| \to 1$ for $n \to \infty$, then also $|a_n| \to 1$ and inversely. The case $|z_n| \to 1$ is certainly an analog of the spiral behavior in the continuous dynamics. In the contrary, if the iterating point remains in the close area of radius less then unit, then it is an analog of the compact behavior. Numerical results give us both types of the behavior. In Fig.5 and Fig.6, $\chi_n$ rotates with $n$ and simultaneously its absolute value changes periodically with iteration.

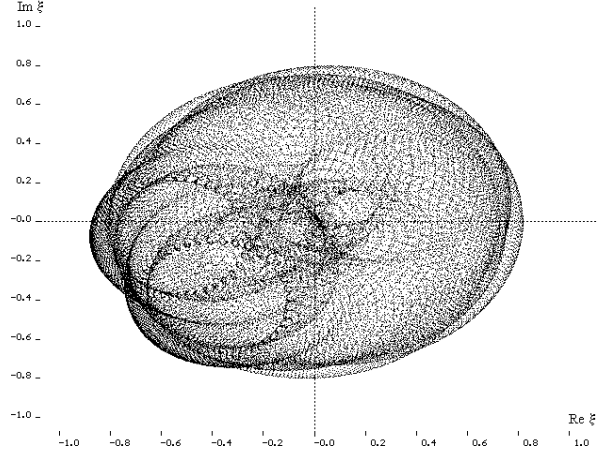

Fig.5: Discrete dynamics for $\chi_n = 0.55\sin(Tn)\exp(iTn)$. Parameters: $T = 100$; $t = 0.1T$; $\omega = 1$; $\xi_0 = 0$.

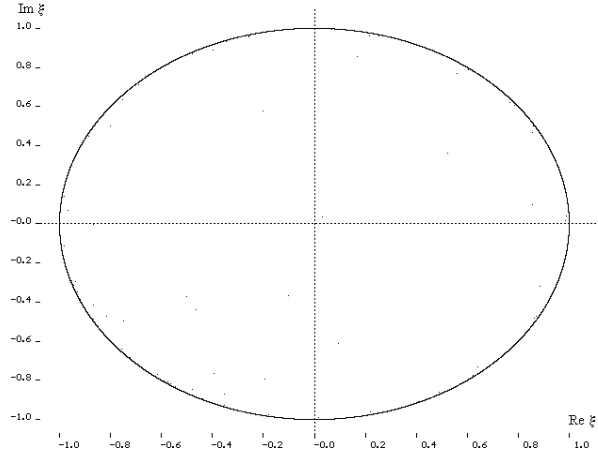

Fig.6: Discrete dynamics for $\chi_n = 0.56\sin(Tn)\exp(iTn)$. Parameters: $T = 100$; $t = 0.1T$; $\omega = 1$; $\xi_0 = 0$.

These pictures illustrate that there exist some critical value of coefficient $\chi_n$ (for given time segments) when behavior drastically changes from compact to spiral behavior.

**4.2 Fixed points**

Fixed points for models, which are special cases of Möbius automorphisms of the unit circle has been already discussed in literature, e.g. [3]. Here we briefly examine the general iteration eq.(28). Its fixed points given by the condition $z_n = z_{n+1}$, fulfill the equation

$$\bar{a}z^2 + \left(e^{i\theta} - 1\right)z - ae^{i\theta} = 0 . \tag{32}$$

where $a, \theta$ stand here for $a^{(n)}, \theta^{(n)}$. The solutions are

$$z = -i\frac{e^{i\theta}}{\bar{a}}\left[\sin\left(\frac{\theta}{2}\right) \pm \sqrt{\sin^2\left(\frac{\theta}{2}\right) - |a|^2}\right] . \tag{33}$$

For $\sin^2(\theta/2) > |a|^2$ there are two elliptic fixed points, inverse to each other with respect to the unit circle. So only one of them lies in interesting us area of the unit circle. In parabolic case, when $\sin^2(\theta/2) = |a|^2$ both points meet each other in the same place of the unit circle. And for $\sin^2(\theta/2) < |a|^2$ there are two hyperbolic fixed points on the unit circle. One of them is attractive and the other is repulsive. In general $\chi$ is iteration-dependent and so are parameters $a^{(n)}, \theta^{(n)}$. Then, the above formula gives points which are fixed only in transition from step *n* to *n* + 1. In fact we can treat eq. (33) as an equation of motion for fixed points. Finally, we explicitly write down the critical equation: $\sin^2(\theta/2) = |a|^2$. For the special model described by eq.(27), it takes the form

$$\{p_n \cos(p_n t)\sin[\omega(T-t)] + \omega \sin(p_n t)\cos[\omega(T-t)]\}^2 - |\chi_n|^2 \sin^2(p_n t) = 0. \quad (34)$$

We plot the right side of the equation as a function of $|\chi_n|$ in the 'physical range' $\omega > |\chi|$, (Fig.7).

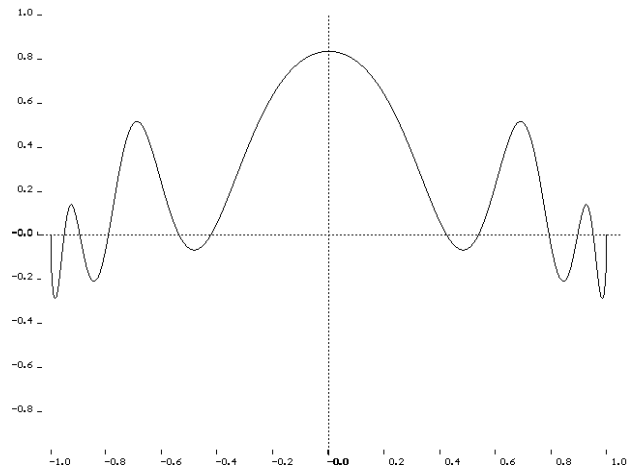

Fig.7: Plot of function $f = f(|\chi_n|)$, (see eq.34), in range $\omega > |\chi_n|$. Parameters $T = 20$; $t = 0.5\,T$; $\omega = 1$.

One of the solutions of eq.(34) is $|\chi_n| = \omega$ independently of parameters $T$, $t$. Note, that in contrary to the continues dynamics, there are ranges, where hyperbolic points appear, and ranges where only elliptic point exists even for the iteration-independent $\chi$ (for given $T$, $t$).

**4.3 Linear transformations**

We show here, that in some sense, we may shift the problem from Möbius mappings of the unit circle to the linear transformations of it. The evolution of the iterated point is ruled by the chain *M* of Möbius mappings (28)

$$M \equiv W_0 \to W_1 \to W_2 \to \ldots \to W_{n-1} \to W_n \quad (35)$$

where each $W_k$, $k = 0, 1, \ldots, n$ means a unit circle. The above chain can be expressed by a new

one $L$

$$M \equiv W_0 \to L \to W_n \quad , \text{ where} \tag{36}$$

$$L \equiv Z_0 \to W_0 \to W_1 \to Z_1 \to W_1 \to W_2 \to Z_2 \to ... \to Z_{k-1} \to W_{k-1} \to W_k \to Z_k \to \tag{37}$$
$$... \to Z_{n-1} \to W_{n-1} \to W_n \to Z_n$$

and transformations $W_k \to Z_k$ and $Z_k \to W_k$ are each other reciprocal. We also assume that $Z_k$ are unit circles and mappings $W_k \to Z_k$ are some Möbius automorphisms. We want to make the transformation

$$Z_{k-1} \to Z_k \equiv Z_{k-1} \to W_{k-1} \to W_k \to Z_k \tag{38}$$

linear, i.e., it should map infinity on infinity. Then $L$ becomes the chain of linear mappings, and linear mappings of a unit circle are simply rotations of the circle.

$$L \equiv Z_0 \to Z_1 \to Z_2 \to ... \to Z_{n-1} \to Z_n \tag{39}$$

The linear mapping $L$ may be easily found. Then using (36) one may find expression for $w_n(w_0)$. Unfortunately, in general it is impossible to accomplish. The exception is, if parameters $a, \theta$ do not depend of iteration, then (in every step) it is possible to make that the first mapping in (right side) (38) maps infinity in fixed point of the second mapping in (38), and then the third mapping in (38) is reciprocal to the first. Using formula (36) we obtain $w_n$ via initial condition $w_0$. It takes on the form of eq. (30) with the following effective parameters

$$a_n = \frac{a}{1 - e^{-i\theta/2}(1-|a|^2)K_n} \quad , \quad e^{i\theta_n} = e^{i\theta}\frac{\bar{a}_n a}{a_n \bar{a}} \tag{40}$$

where $\quad K_n = \frac{(+)^n - (-)^n}{(+)^{n+1} - (-)^{n+1}} \quad$ and $\quad \begin{aligned}(+) &= \cos(\theta/2) + \sqrt{|a|^2 - \sin^2(\theta/2)} \\ (-) &= \cos(\theta/2) - \sqrt{|a|^2 - \sin^2(\theta/2)}\end{aligned} \tag{41}$

The above result can be obtained in many different ways. In the special case of kicked dynamics it was given in [3] but without prove.

## 5. FINAL REMARKS

In the first part of the paper some features of the continues dynamics of SU(1,1) have been described. We find two distinctive types of behavior. The first one where trajectories tend to the unit circle as a limit (spiral behavior) and the second one, where trajectories form more or less complicated figures inside unit circle and do not tend to the limit (compact behavior).

In the next part of the paper we have shown that it is convenient to examine all pulsed SU(1,1) group models from the general point of view of Möbius automorphism of the unit circle. The two distinctive types of behavior occurring in continuous case can also be identified in pulsing cases. As an example of a pulsed system serves us the model of finite width and magnitude of pulsing peak. Some numerically obtained pictures, e.g. Fig.5, look like 'chaotic'. Nevertheless it can be easily proved that the motion is in fact regular, because the Lapunov exponent is either zero or negative. In the latter case it may be finite or infinite. The zero Lapunov exponent corresponds to the compact behavior. The negative Lapunov exponents correspond to spirals. There is no positive Lapunov exponent and it means there is no classical chaos in the system. There were some attempts to look for 'fingerprints' of quantum chaos in SU(1,1) systems, because it seems that the quantum autocorrelation function exhibits some decay, [6] (see also [10]). However that decay appears when the type of motion changes from the dynamics described by zero Lapunov exponent to the dynamics described by negative Lapunov exponent (compare Fig.5 and 6). The classical and quantum mappings based on the Hamiltonian (1) are equivalent. We think it is rather strange to look for 'fingerprints' of quantum chaos, when the classical counterpart becomes even more regular than before transition.

## ACKNOWLEDGMENTS

I would like to thank Prof. A. Bechler for many useful discussions during preparation of the paper. Without his assistance this paper would never have come into existence.